\documentclass[sigconf]{acmart}

\usepackage{booktabs} 
\DeclareMathAlphabet{\mathpzc}{OT1}{pzc}{m}{it}
\usepackage{algorithm}
\usepackage{mathtools,xparse}
\usepackage{amsmath,amssymb}
\usepackage{flexisym}
\usepackage[english]{babel}
\usepackage[utf8]{inputenc}
\usepackage{listings}
\usepackage[noend]{algpseudocode}
\usepackage{booktabs}
\usepackage{bbold}
\usepackage{tikz}
\usepackage{array}
\usepackage{makecell}
\usepackage{tabularx,ragged2e}
\usepackage{svg}
\usepackage{balance}
\DeclarePairedDelimiter{\norm}{\lVert}{\rVert}
\NewDocumentCommand{\normL}{ s O{} m }{%
  \IfBooleanTF{#1}{\norm*{#3}}{\norm[#2]{#3}}_{L_2(\Omega)}%
}

{
\theoremstyle{plain}

}

\acmArticle{4}

\begin{document}
\title{Conversational Recommender System}

\author{Yueming Sun, Yi Zhang}
\affiliation{%
  \institution{University of California, Santa Cruz}
  \streetaddress{1156 High St.}
  \city{Santa Cruz} 
  \state{CA} 
  \postcode{95064}
}
\email{yueming, yiz@soe.ucsc.edu}

\begin{abstract}

A personalized conversational sales agent could have much commercial potential. E-commerce companies such as Amazon, eBay, JD, Alibaba etc. are piloting such kind of agents with their users. However, the research on this topic is very limited and existing solutions are either based on single round adhoc search engine or traditional multi round dialog system. They usually only utilize user inputs in the current session, ignoring users' long term preferences. On the other hand, it is well known that sales conversion rate can be greatly improved based on recommender systems, which learn user preferences based on past purchasing behavior and optimize business oriented metrics such as conversion rate or expected revenue. In this work, we propose to integrate research in dialog systems and recommender systems into a novel and unified deep reinforcement learning framework to build a personalized conversational recommendation agent that optimizes a per session based utility function. 

In particular, we propose to represent a user conversation history as a semi-structured user query with facet-value pairs. This query is generated and updated by belief tracker that analyzes natural language utterances of user at each step. We propose a set of machine actions tailored for recommendation agents and train a deep policy network to decide which action (i.e. asking for the value of a facet or making a recommendation) the agent should take at each step. We train a personalized recommendation model that uses both the user's past ratings and user query collected in the current conversational session when making rating predictions and generating recommendations. Such a conversational system often tries to collect user preferences by asking questions. Once enough user preference is collected, it makes personalized recommendations to the user. We perform both simulation experiments and real online user studies to demonstrate the effectiveness of the proposed framework. 
\end{abstract}

%
%



\copyrightyear{2018} 
\acmYear{2018} 
\setcopyright{acmlicensed}
\acmConference[SIGIR '18]{The 41st International ACM SIGIR Conference on Research \& Development in Information Retrieval}{July 8--12, 2018}{Ann Arbor, MI, USA}
\acmBooktitle{SIGIR '18: The 41st International ACM SIGIR Conference on Research \& Development in Information Retrieval, July 8--12, 2018, Ann Arbor, MI, USA}
\acmPrice{15.00}
\acmDOI{10.1145/3209978.3210002}
\acmISBN{978-1-4503-5657-2/18/07}

\maketitle

\section{Introduction}

As intelligent assistants such as Siri (Apple), Facebook Messenger, Amazon Alexa, Google Assistant, enter the daily life of users, research on conversational information systems is becoming increasingly important. There are mainly three kinds of conversational systems (i.e. dialogue systems): chit-chat, informational chat and task oriented chat. Chit-chat systems are focusing on information social chat and try to interact with human-like reasonable or interesting responses \cite{li2015diversity}\cite{vinyals2015neural}. Informational chatbots try to help user find information or directly answer user questions. Task oriented chatbots try to help users finish a specific task, such as booking a flight or canceling a trip. And they are usually built for a close domain. This paper is related to both informational chat and task oriented chat.

Due to the big commercial potential, there are quite some activities on task oriented conversational chatbots that can interact with users to help them find products/services. Companies like Amazon, Google, eBay, Alibaba are all rolling out these chatbots. Most of existing works are focusing on natural language processing or semantic rich search solutions for dialogue systems. The most notable recent related work is \cite{dhingra2017towards}, which focuses on enabling user to query knowledge base interactively. 

On the other hand, researchers have demonstrated the importance of recommender systems in e-commerce websites and applications. To improve the success or conversion rate of a shopping/sales chatbot, we argue that one should integrate recommendation techniques into conversational systems. Intuitively, this can benefit both recommender systems and dialog systems. For dialogue systems, good recommendations based on users' previous purchasing or rating history can better fulfill user's information need, and create more business opportunities. For recommender systems, dialogue systems can provide more detailed information about user intentions, such as user preferred price range or the location of a restaurant, by interactively soliciting and identifying user intentions based on multi-round natural language conversation. This motivates us to study how to build a conversational recommender system.

This paper tries to integrate search and recommendation techniques with conversational systems seamlessly. We build a chat agent that can assist users to find items interactively. With the recent breakthrough of deep learning technologies and a better understanding of search and recommendation, we can approach this problem with a new perspective and a set of enabling technologies. Similar to other dialog systems, our system has three major components. First, a natural language understanding (NLU) module for analyzing each user utterance, keeping track of the user's dialogue history and constantly updating the user's intention. This NLU module focuses on extracting item specific meta data. Second, we propose a dialogue management (DM) module that decides which action to take given the current state. This DM module has an action space defined specifically for this task. It is well integrated with an external recommender system. The third component is a natural language generation module to generate response to the user. This framework enables us to build a conversational search and recommender system that can decide when and how to gather information from users and make recommendations based on a user's past purchasing history and context information in the current session. 

For the NLU module, we train a deep belief tracker to analyze a user's current utterance based on context and extract the facet values of the targeted item from the user utterance. Its output is used to update the current user intention, which is represented as a user query that is a set of facet-value pairs about the target. The user query will be used by both the dialogue manager and the recommender system. For the DM module, we train a deep policy network that decides which machine action to take at each turn given the current user query and long term user preferences learned by the recommender system. The action could be asking the user for information about a particular facet or recommending a list of products. The deep policy network selects an action that maximizes the expected reward in the entire conversation session. When the user query collected so far is sufficient to identify the user's information need, the optimal action usually is recommending a list of items that is personalized for the user. When the user query collected is not sufficient, the optimal action usually is asking for more information.

\section{Related Work}

There are four lines of research that are closely related to and motivate our work: conversational dialogue system, recommender system, faceted search and the deep reinforcement learning.

\subsection{Dialogue System}

There have been three main streams of dialogue systems (DS): the chit-chat DS, informational DS and task-oriented DS. Early works of task oriented DS require large amount of labeled data \cite{young2013pomdp}\cite{williams2013dialog} and are very expensive. Recent works tend to apply deep learning techniques on each component of the dialogue system and have demonstrated significant improvements. Mesnil et al. \cite{mesnil2015using} used LSTM and Conditional Random Fields networks to perform the slot filling. Wu et al. \cite{wu2015probabilistic} developed an entropy based policy for the DS. Christakopoulou et al. \cite{christakopoulou2016towards} used bandit machine for the decision making. Zhao et al. \cite{zhao2016towards} was among the first works of building an end-to-end dialogue system. 
Wen et al. \cite{wen2016network} introduced an end-to-end task oriented dialogue system and a wizard-of-oz framework for data collection in the restaurant domain. Bordes et al. \cite{bordes2016learning} built an end-to-end task oriented bot based on memory network. Dhingra et al. \cite{dhingra2017towards} built a goal oriented information access system based on reinforcement learning, trying to select related items with certain attribute values. However, most of those prior works focus on NLP challenges instead of commercial success metrics such as conversion rate. They either did not focusing on recommendation problems or did not model or utilize the user's past preferences when recommending items to users.

\subsection{Recommender Systems}

Recommender systems have achieved much commercial success and are becoming increasingly popular in a wide variety of practical applications. For example, online stores such as Amazon, iTunes and Netflix provide customized recommendations for additional products or services based on a user's history. Most recommender systems are either content based \cite{lops2011content}, collaborative filtering (CF) based \cite{koren2009matrix}\cite{mnih2008probabilistic} or hybrid. 
Research in recommender systems usually focus on improving rating prediction or ranking measures (learning to rank for recommendation) \cite{koren2009matrix}\cite{rendle2010factorization}. 
Few work has been done towards making recommendations in a dialogue system. Christakopoulou et al. \cite{christakopoulou2016towards} studied using a generative Gaussian model to recommend items to users in a conversation. However, their model does not target at maximizing the long-term benefits and their conversational agents' action space is very limited and doesn't include actions one would typically expect in a dialog system. Their dialog system only asks questions about whether a user likes an item or whether the user prefers item A to item B, while a typical task oriented dialogue system often solicits facets from users \cite{wen2016network}\cite{dhingra2017towards}. In this work, we maximize the long-term utility by using the reinforcement learning framework, and our question types are requesting facets from users and recommending a list of items to users, which are better aligned with the typical dialogue systems.

\subsection{Faceted Search} 

Conversational recommendation agent interactively helps a user find candidate items. Faceted search, a commonly used web technique in the e-commerce domain, is also a technique that interactively helps a user find candidate items. On faceted-search-enabled websites, buyers can narrow down their list of products by adding constraints on a group of merchandise facets \cite{vandic2017dynamic}. It has been shown that a well designed faceted search idea can be understood by the average user. Users might have preferences for certain types of facets. Movie viewers might have preferences on movie genres, directors or actors; shoe buyers might have preferences on brands and colors, while restaurant seekers might prefer a region of food or a price range, etc. To avoid overwhelming users with too many facet-value pair options per conversation, a faceted search engine selects a small set of facets or facet-value pairs for a user to choose from based on context \cite{zhang2010interactive}\cite{koren2008personalized}. Motivated by prior work, we introduce a particular type of machine actions into our conversational system: selecting a facet based on the context and asking user to provide information about her preferred facet value, such as "What's the color you like?", "Which brand you prefer?", "Do you like small size, middle size or large size?". Unlike prior work on faceted search, our facet selection decision is made by a deep reinforcement learning algorithm.

\subsection{Deep Reinforcement Learning}

Deep learning techniques allow people to use deep neural networks for function approximation in reinforcement learning (RL) \cite{sutton1998reinforcement}. One of the most famous success of deep RL is Google's DeepMind research on the game of Go \cite{silver2016mastering}\cite{silver2017mastering}. Deep RL has been applied for better sequential decision making in various domains, including End-to-End dialogue systems using deep RL for information access \cite{dhingra2017towards} \cite{zhao2016towards},  information extraction \cite{narasimhan2016improving}, query reformulation \cite{nogueira2017task}, real time ads bidding \cite{cai2017real}. Shani et al. \cite{shani2005mdp} is one of the early studies of applying RL techniques for the recommender system problems, however, not in dialogue systems. Inspired by these works, we build a deep RL based conversational recommender system. This combines the ranking and personalization ability of recommender system with the sequential decision making power of the RL models, thus can better serve a user.

\section{Conversational Recommendation with Reinforcement Learning}

\subsection{Overview}

In this section, we discuss how to tackle the problem of building conversational recommender system. Our framework has three components: a belief tracker, a recommender system and a policy network.

\begin{figure}[!ht]
\centering
\begin{minipage}{\linewidth}
\centering
\includegraphics[width=1\linewidth]{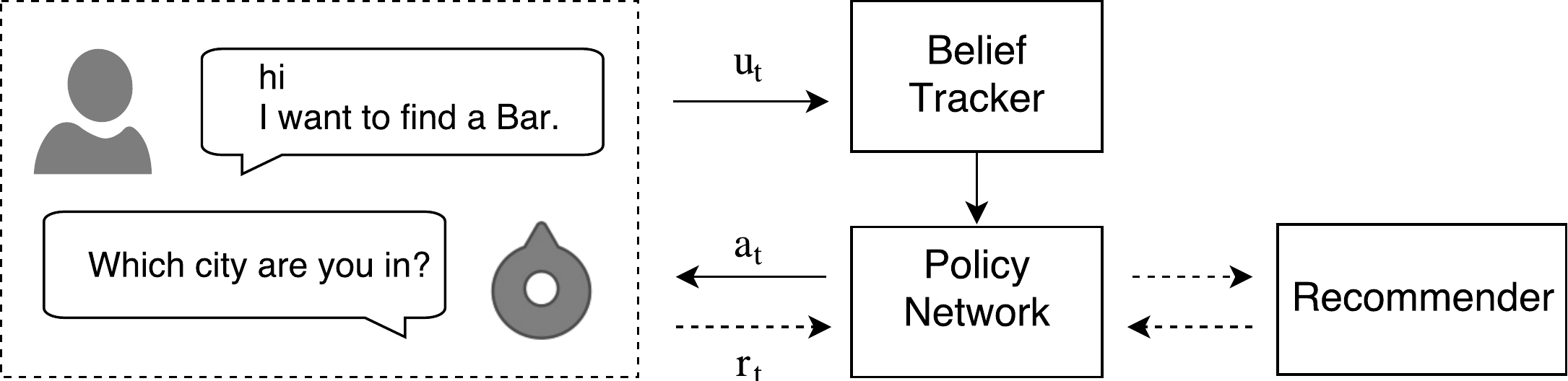}
\footnotesize
\caption{The conversational recommender system overview}
\label{fig:overview}
\end{minipage}
\end{figure}

The goal of our conversational recommender system is successfully recommending good item(s) to a user, and it achieves this goal by analyzing what the user has said in the current session, interactively asking user clarification questions, and making personalized recommendations when appropriate, based on the current session and what the user has consumed or rated before. 

Several aspects are important in the process. First, how to understand the user's intention correctly. Second, how to make sequential decisions and take appropriate actions in each turn. Third, how to make personalized recommendations in order to maximize the user satisfaction. Figure~\ref{fig:overview} presents the overview of our proposed framework. At a time step in the dialogue, the user utters ``I want to find a Bar''. The framework calls the belief tracker to convert the utterance into a vector representation or ``belief''; then the belief is sent to the policy network to make a decision. For example, the policy network may decide to request the city information next. Then the agent may respond with ``Which city are you in?'', and gets a reward, which is used to train the policy. A different decision is to make a recommendation. Then the agent calls the recommender system to get a list of items personalized for the user. We introduce each component and the relationships among them in more details in the following sections.

\begin{figure}[!ht]
\centering
\begin{minipage}{\linewidth}
\centering
\includegraphics[width=1\linewidth]{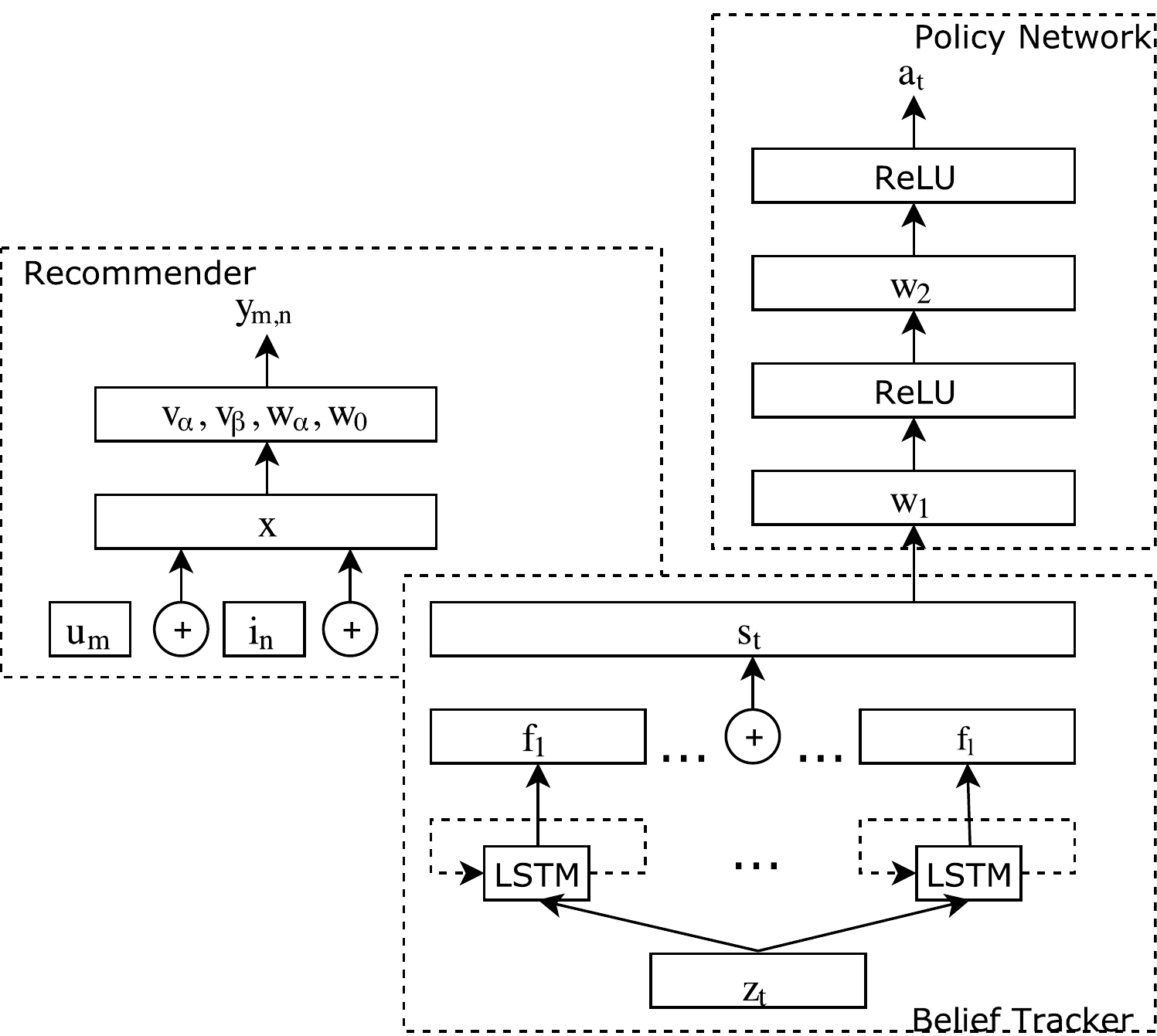}
\footnotesize
\caption{The structure of the proposed conversational recommender model. The bottom part is the belief tracker, the top left part is the recommendation model, and the top right part is the deep policy network.}
\label{fig:structure}
\end{minipage}
\end{figure}

\subsection{Belief Tracker}

When trying to buy products on an e-Commerce website, users often navigate the product space through faceted search \cite{zhang2010interactive}\cite{koren2008personalized}\cite{vandic2017dynamic}. 
Motivated by this and in order to assist users to find the item they want in conversation, it is crucial that the system understands which values the user has provided for product facets, and represents the user utterances with a semi-structured query. We introduce a Belief Tracker module similar to \cite{dhingra2017towards} to extract facet-value pairs from user utterances during the conversation, and maintain the facet-value pairs as the memory state (i.e. user query) of the agent. In this paper, we view the product facet (or attribute, metadata) $f$ along with its specific value $v$ as a facet-value pair $(f, v)$. Each facet-value pair represents a constraint on the items. For example, $(color, red)$ is a facet-value pair which constrains that the items need to be red in color.

The network structure of belief tracker is shown in the lower part in Figure~\ref{fig:structure}. We train a belief tracker for each facet of the items. The belief tracker takes the current and the past user utterances as the input, and outputs a probability distribution across all the possible values of a facet at the current time point. The dialogue system's belief of the session is constituted by the predicted values of different facets. Specifically, given a user utterance at time step t, $e_t$, the input to the belief tracker is the n-gram vector $\textbf{z}_t$, and the dimension of $z_t$ is the size of the n-gram vocabulary.

\begin{equation}
\textbf{z}_t = ngram(e_t)
\end{equation}

Next, the sequence of n-grams up to the current time is encoded by a LSTM network into a vector $\textbf{h}_t$, which is then fed to a softmax activation layer to be transformed to a probability distribution across the available values. The softmax layer's output size is $V^j$, for a categorical facet holding $j$ possible values.

\begin{equation}
\textbf{h}_t = LSTM(\textbf{z}_1, \textbf{z}_2, ..., \textbf{z}_t)
\end{equation}
\begin{equation}
\textbf{f}_{i} = softmax(\textbf{h}_t)
\end{equation}

At each round, all the $\textbf{f}_i$ are concatenated to each other to form the agent's current belief of the dialogue state in the current session. If there are $l$ facets, then

\begin{equation}
s_t = \textbf{f}_1 \oplus \textbf{f}_2 ... \oplus \textbf{f}_l
\end{equation}

where $\textbf{f}_i$ is the learned vector representation for the facet $i$, $i\leq l$. By using the learned output of the LSTM network directly, we keep the uncertainty from the belief tracker for the following modules.

\subsection{Recommender System}

As the conversational system interacts with the users, at certain round, the conversational system can decide to make a recommendation based on its current belief of the user's information need, which is interpreted as the dialogue state. We train the recommender with the dialogue state, user information and item information. Specifically, we use the Factorization Machine (FM) \cite{rendle2010factorization}, for the reason that FM can combine different features, e.g. $s_t$, together to train the recommendation model.

The structure of recommendation model is shown in the upper left part of Figure~\ref{fig:structure}. Let \textbf{U} denote the users and \textbf{I} the items. For $M$ users and $N$ items in the dataset, the users and items are represented as the sets: $\{ u_1, u_2, ..., u_M\}$ and $ \{ i_1, i_2, ..., i_N\}$. The input feature $\textbf{x}$ is the concatenation of the 1-hot encoded user/item vector, where the only element that is not zero in the vector corresponds to the index of the encoded info, and the dialogue belief:

\begin{equation}
\textbf{x} = \textbf{u}_m \oplus \textbf{i}_n \oplus \textbf{s}_t 
\end{equation}

\begin{equation}
\textbf{u}_m = \{0, 0, ..., 1, ..., 0\}, with \; 1 \; at \; the \; m_{th} \; element.
\end{equation}

\begin{equation}
\textbf{i}_n = \{0, 0, ..., 1, ..., 0\}, with \; 1 \; at \; the \; n_{th} \; element.
\end{equation}

{\setlength{\parindent}{0cm}
where $m$ and $n$ denotes that $i_n$ is rated by the $u_m$. The output $\textbf{y}_{m,n}$ can be either a rating score for the explicit feedback or a 0-1 scalar for the implicit feedback. We use a 2-way $(K=2)$ FM:
}

\begin{equation}
\textbf{y}_{m,n} = \textbf{w}_0 + \sum_{\alpha=1}^N \textbf{w}_{\alpha} \textbf{x}_{\alpha} + \sum_{\alpha=1}^N \sum_{\beta=\alpha+1}^N \langle \textbf{v}_{\alpha}, \textbf{v}_{\beta}\rangle \textbf{x}_{\alpha}, \textbf{x}_{\beta}\\
\end{equation}

\begin{equation}
\langle \textbf{v}_{\alpha}, \textbf{v}_{\beta}\rangle = \sum_{\kappa=1}^{K} \textbf{v}_{\alpha,\kappa} \textbf{v}_{\kappa,\beta}
\end{equation}

{\setlength{\parindent}{0cm}
where $\textbf{w}_0$, $\textbf{w}_{\alpha}$, $\textbf{v}_{\alpha}$ and $\textbf{v}_{\beta}$ are learnable parameters. $\alpha$ and $\beta$ denote the index of the input vector $\textbf{x}$, and $\textbf{y}_{m,n}$ is the $u_m$'s feedback to $i_n$. For rating prediction, stochastic gradient descent is used to minimize the L2 loss between the predicted rating score and the real rating score. The objective function scales linearly with the size of the data.
}

Without loss of the generality, at the time of making recommendations using the trained model, we first take the $argmax$ of each facet's belief, to get $l$ categorical distributions over the values, one for each facet. The combinations of the facet values form a new distribution, with the probability the product of $l$ value's probabilities. Then we keep the $\mu$ most probable combinations, and use their facet values to retrieve items from the entire item set. The retrieved items form a candidate set. Then we use the trained model to re-rank the candidates based on their rating scores.

\subsection{Deep Policy Network}

We now describe the deep policy network we use to manage the conversational system. At each turn, the reinforcement learning model selects an action based on the dialogue state, in order to maximize the long term return. Detailed introduction of reinforcement learning can be found at \cite{sutton1998reinforcement}. Here we adopt the policy gradient method of reinforcement learning, which can learn a policy directly, without consulting the value functions \cite{sutton1998reinforcement}. 

The structure of the policy network is shown in the upper right part in Figure~\ref{fig:structure}. The reinforcement learning has the basic components of state $S$, action $A$, reward $R$ and policy $\pi(a|s)$.

\textbf{State: } The state $s_t$ is the current description of the environment from the viewpoint of the agent. In our case, it is the description of the conversation context, which is the belief tracker's output, $s_t = \{\textbf{f}_1 \oplus \textbf{f}_2 ... \oplus \textbf{f}_l\}$.

\textbf{Action: } An action $a_t$ is the decision the agent needs to make at time step $t$. Here we have mainly two kinds of actions. One is to request the value of a facet, which is further divided into $l$ actions $\{a_1, a_2, ..., a_l\}$, one per each facet. The other is to make a personalized recommendation $a_{rec}$, in which case the recommendation module described above would be called. Note that $a_{rec}$ may in fact occur more than once in a single conversation session. We leave the modeling of multiple recommendations to the future work.

\textbf{Reward: } The reward is the benefit/penalty the agent gets from interacting with its environment. At each turn, according to the current state $s_t$, the agent selects an action $a_t$ following the policy, and it gets an immediate reward $r_t$, denoting how good the current decision is. The state $s_t$ transits to a new state $s\textprime$. In our case, the conversational system gets a reward when it requests a facet value, or makes a recommendation. Note that the reward is the feedback from the environment. Our recommender system serves as part of the environment. The agent only gets rewards from the environment but can not change it \cite{sutton1998reinforcement}. We model the recommendation reward in different ways, which will be introduced in section \ref{recommendation_rewards}.

\textbf{Policy: } This is the target the model tries to learn. Usually denoted as $\pi(a_t|s_t)$, the policy represents the score, such as the probability, of taking action $a_t$ when the agent is in state $s_t$. For simplicity and without loss of generality, we use two fully connected layers as the policy network, each layer with a $ReLU$ activation function. Other deep neural network structures may also work. The output of the network is further sent to a softmax layer. Specifically, the goal of the policy network is to maximize the episodic expected reward from the starting state: 

\begin{equation}
\eta(\theta)=E_\pi[\sum_{t=0}^{T}\gamma^t r_t ]
\end{equation}

where $\theta$ is the policy parameter to be learned, $\gamma$ is a discount parameter emphasizing more on the immediate reward than the future rewards, and $r_t$ is the reward at time step $t$. Following the REINFORCE \cite{williams1992simple} algorithm, the gradient of the learning object becomes:

\begin{equation}
\nabla \eta(\theta) = E_\pi [\gamma^t G_t \nabla_\theta \log \pi(a_t | s_t, \theta) ]
\end{equation}

Here $G_t$ is the sum of rewards, or \textbf{return}, starting from time step $t$ to the final time step $T$. Note that in our case we always have a terminating state of the conversation, e.g., the user leaves the chat or the user is successfully recommended with a target. Each such kind of sequence is called an episode in the reinforcement learning literature \cite{sutton1998reinforcement}. This enables us to use gradient descent methods to optimize the parameter $\theta$ in the policy network directly. Note that if $\theta$ is initialized randomly, the learning can fail completely. To address this issue, we use a rule based policy, which is introduced in the next section, to initialize the parameters.

\section{Experimental Setup}

We conduct both offline experiments with simulated user and online experiments with real users to study multi round conversational recommendation agents proposed.

\subsection{Dataset}

To build the proposed system, we adapt the restaurants and food data of the yelp challenge recommendation dataset\footnote{https://www.yelp.com/academic_dataset} to create the data needed for our study. To generate dialogue scripts, five item attributes are selected as the candidate facets. Users and the items that have less than 5 reviews are removed. The statistics of the dataset are shown in the Table~\ref{tab:data_statistics}. The facet values for category include Mexican, Mediterranean, etc. The rating includes 5.0, 4.5, 4.0, etc.

\begin{table}
  \caption{Basic data statistics}
  \label{tab:data_statistics}
  \begin{tabular}{ l | c }
    \hline
   	\hline
      & Number of Values \\ \hline
    \hline
    Users & 62047 \\ \hline
    Items & 21350 \\ \hline
    User-item pairs & 875721 \\ \hline
    Category & 191 \\ \hline
    State & 13 \\ \hline
    City & 189 \\ \hline
    Price & 4 \\ \hline
    Rating Range & 9 \\ \hline
  \end{tabular}
\end{table}

\subsection{User Simulation}

The reinforcement learning agent always needs an environment to interact with. For training games like Go or Atari \cite{silver2017mastering}\cite{mnih2013playing}, such kind of environment is easy to create based on a set of predefined rules. However, for dialogue system, the environment needs ``real users'' to chat with the agent and provide rewards. It's hard to create such kind of environment for research. Even for companies with millions users, launching a dialogue system without much training to real users is likely to fail due to the poor initialization parameters and poor performance \cite{dhingra2017towards}.

To tackle this problem, a common practice in the dialogue system research is using simulated users as a bootstrap to pre-train the model \cite{bordes2016learning}\cite{li2017investigation}\cite{thomson2012n}. Without lose of generality and based on (user, target item) yelp data, we created simulated users following a very simple agenda to interact with the agent. The simulated users are used to pre-train the agents. Then we conduct both offline experiments with simulated users and online experiments with real users to show the effectiveness of the learned agents. 

The goal of the simulated user is to chat with the conversational system to find the target item. She first informs the facet values of a target item to the agent. When the agent understands and makes a recommendation, the user would examine the list to try to find the target item. A user has the following 3 behaviors: 1) answering the agent's question. When the chatbot asks for the value of a new facet, the user would respond with natural language containing its value; 2) finding the target item in the recommendation list. When the chatbot recommends a list of items, the user would ``view'' the items one by one, until it finds the item (success) or fails; 3) leaving the dialogue. The user leaves the dialogue if the dialogue is too long, or if the target item is not in the recommendation list, or the target item is in the list but is ranked too low.

For each behavior, the user gives a numerical reward to the agent. Formally, let $r_q$ denote a negative reward when the user quits the conversation; $r_p$ denote the positive reward when the user successfully finds the target in the recommendations; and $r_c$ be a small negative reward per dialogue turn, preventing the dialogue from getting too long. Note that once a set of rewards is fixed, the environment setting is fixed, then the optimal policy is also fixed. Algorithm~\ref{alg:framework} summarizes the interaction.

\begin{algorithm}
\caption{The interaction between agent and simulated user}
\label{alg:framework}
\begin{algorithmic}[1]
\State Start with M epochs, N training data
\For{epoch = 1, M}
  \For{i = 1, N}
    \State $t$ = 0
    \State Sample a $(u, i)$ pair from the training set
    \State The user $u$ starts the conversation, and conveys a random facet value with utterance $e_t$
    \While{True}
      \State Apply belief tracker to $e_t$ to get $s_t$
      \State Send $s_t$ to the policy gradient agent
      \State Get action $a_t$ from the agent
        \If{$a_t$ is $a_{rec}$}
          \State Call the recommender to get an item ranking list 
          \If {the target item is in the top K}
            \State $r_t$ = $r_p$, the dialogue succeeds and $break$
          \Else 
            \State $r_t$ = $r_q$, the dialogue fails and $break$
          \EndIf
        \Else 
          \State The system generates a response $e_m$ to the user
          \If {the user quits}
            \State $r_t$ = $r_q$, the dialogue fails and $break$
          \Else
            \State $r_t$ = $r_c$, t = t + 1
            \State The user responds with a new utterance $e_{t}$
          \EndIf
        \EndIf
    \EndWhile
  \EndFor
\EndFor
\end{algorithmic}
\end{algorithm}

\subsection{Recommendation Rewards} 
\label{recommendation_rewards}

Let's assume a user stops checking after seeing $K$ items, where $K$ is a threshold. Let C denote the maximum reward one can get when the target is ranked to the first position. 

When recommended with a list of items, a user may review the list differently. Therefore we model the success reward $r_p$ in the following different ways. 

\textbf{Linear $r_p$: } This is the most straight forward way. We assume the user always checks the next item with probability $1$ until she either finds the target or reaches the threshold. $r_p=\frac{C * (K - \tau + 1)}{K}$ when $\tau \leq K$, where $\tau$ is the ranking of the target. If $\tau > K$, this is denoted as a failure.

\textbf{NDCG $r_p$: } The second way is following the assumption of NDCG \cite{Manning:2008:IIR:1394399}. In this case we assume the items ranked higher are more preferred than the items ranked lower. When computing the NDCG, we use a binary relevance score. So $r_p = C * NDCG@K$, if a user finds the target.

\textbf{Cascade $r_p$: }The third way is based on the cascade model \cite{craswell2008experimental}. The chatbot has a limited UI space and can only show $\kappa$ recommended items. We assume a user can horizontally scroll to the next page to view more items page by page. Within each page, the user views all items. For each page, the user has a probability of $p_r$ to continue and $1-p_r$ to leave, with $p_r$ decaying exponentially with a factor $\alpha_1$. The reward is also decaying exponentially with a factor $\alpha_2$, where $0 < \alpha_1, \alpha_2 < 1$. Thus we have $r_p=C * \alpha_2^{\rho - 1}$, where $\rho$ denotes the page number with $\rho \leq \lceil K/\kappa \rceil$. In this paper we set $\kappa=3$.

\subsection{User Utterance Collection}

First of all, we use the collected data to pretrain our belief tracker and to perform the offline experiments. However for the online experiments, our trained model is interacting with the real user's natural language utterances. By using a user's historical ratings, the system can chat with users and make recommendations to the user at the same time. 

The yelp recommendation dataset contains rich rating information, however, it doesn't include any dialogue utterances. To the best of our knowledge, there is no such available dataset. Thus we created a dataset using Amazon Mechanical Turks. We make a strong assumption that, the users were visiting the restaurants AFTER chatting with a virtual agent. This assumption ensures that we could study the conversational recommendation problems by leveraging the conversational utterances and the rating scores of the user-item pairs in one framework.

Based on this assumption, we create a crowd sourcing task to use a schema based method to collect the dialogue utterances. We assume that the users are cooperative and always inform the facet values of the target.

To collect user utterances, first we sample a target restaurant from the dataset. Then we use a set of templates to generate a dialogue ``schema''. We show this schema to the crowd sourcing workers and ask them to write natural languages that conforms to the templates to complete the dialogue. One example is shown in the Table~\ref{tab:dialogue_schema}. The templates \cite{henderson2013dialog} include: \textit{inform(facet=``value'')}, \textit{recommend()}, \textit{dontknow(facet)} and \textit{thanks()}.

\begin{table}
\small
  \caption{The schema based dialogue collection example. The bold utterances are written by the crowd sourcing workers via rewriting the templates to interact with the agent.}
  \label{tab:dialogue_schema}
  \begin{tabular}{ l }
    \hline
    \hline
    The target restaurant has the following facets. \\
    \{category: Mexican, state: AZ, city: Glendale, \\
    price range: cheap, rating range: >=3.5\} \\ \hline
    \hline
    \makecell{
      \underline{User}: inform(city="Glendale", category="Mexican")\\
      \underline{User Write}: \textbf{I'm looking for Mexican food in Glendale.}\\
      \hspace*{\fill}\underline{Agent}: Which state are you in? \\
      \underline{User}: inform(state="AZ")\\
      \underline{User Write}: \textbf{I'm in Arizona.}\\
      \hspace*{\fill}\underline{Agent}: Which price range do you like?\\
      \underline{User}: inform(price_range="cheap")\\
      \underline{User Write}: \textbf{Low price.}\\
      \hspace*{\fill}\underline{Agent}: What rating range do you want?\\
      \underline{User}: inform(rating_range>="3.5")\\
      \underline{User Write}: \textbf{3.5 or higher}\\
      \hspace*{\fill}\underline{Agent}: <make recommendations>\\
      \underline{User}: thanks()\\
      \underline{User Write}: \textbf{thank you}\\
      } \\ \hline
    \hline
  \end{tabular}
\end{table}

After collecting the utterances, we match the target values back to the written utterances, and substitute them with the placeholders, e.g. <Category>, which is called delexicalization by \cite{wen2016network}. This way we convert the collected user utterances into templates. For instance, ``I'm looking for Mexican food in Glendale'' is converted to the template: ``I'm looking for <Category> in <City>.'' Then we could use these templates to simulate many more dialogues. In total we have collected $385$ dialogues. And by delexicalization we simulated $875721$ dialogues, one for each user-item pair in the yelp dataset. We use these dialogues to train our belief tracker, recommender and the policy network.

\subsection{Baselines}

In order to evaluate our model, we use a Maximum Entropy rule based method as the baseline. This method computes the entropy for each unknown facet, and selects the facet having the maximum entropy to be the next one to ask. 
It stops asking facets when either there is no item meets the current dialogue belief or all facets are known or the dialogue length exceeds the maximum limit of the dialogue turn. Therefore it is a greedy method. This baseline is a rule based policy, because the agent always asks the facets one by one, until it has collected the values of all facets, in which case it makes recommendaitons to users. We further explore its variations that always ask exactly K slots ($K < 5$) before making a recommendation. We name the baseline ``MaxEnt Full'' for the one that asking all the facets and ``MaxEnt@K'' for ones that asking exactly K facets.

\subsection{Evaluation Methodology}

For conversational system, there are three  commonly used measures. The first one is the Average Reward, which measures the long-term gains of the system, defined as the mean of the returns, e.g., $R=\overline{G}_t$. In reinforcement learning, this is the direct learning object and is a major indicator of the effectiveness of RL methods. The second one is the Success Rate. From the interactive dialogue system's perspective, this can be viewed as the conversion rate, which is one of the major measurement a business tries to maximize. The success rate is defined as $S = \frac{\# successful\;dialogues}{\#dialogues} \cdot 100\%$. The third measure is the Average \# of Turns, which is defined as $T=\overline{dialogue\;length}$. In our work, one pair of user-agent utterances is denoted as one turn. Shorter turn indicates that the system can meet the users information need faster, which is better. We report two more measures, the Wrong Quit Rate and the Low Rank Rate. The former shows the rate of failures, when the target item is not included in the recommendation candidate lists due to belief tracker's error. The latter for the failure cases when the target item is ranked too low (lower than the stop threshold).

\subsection{Model Training}

For simplicity and scalability, we use a bag of 2-gram representation for the collected user utterances, and the vocabulary size is $19644$. To train the belief tracker, we split the entire dataset to a train, dev and test set with the $80\%$, $10\%$ and $10\%$ split. We use gradient descent with a learning rate of $0.001$ to train the belief tracker. Then we fix the belief tracker's parameters to train the FM recommender and the policy network. We use the 2-way FM model, with the Adam optimizer and the learning rate of $0.001$. For the policy gradient network, we pre-train it as a classifier, by taking the dialogue state as input and the MaxEnt Full model's actions as the labels. After the classifier's accuracy is stable, we keep on training the policy network using the REINFORCE algorithm \cite{williams1992simple}. We randomly sample $35000$ training dialogues and $2500$ dev and test dialogues from the train, dev and test sets, respectively. We use the RMSProp optimizer with a learning rate $0.001$ and the batch size of $100$. The best belief tracker that we use has an accuracy of $87.5\%$. For the reward signals, unless specifically pointed out, we train the simulated users with Linear $r_p$. We set $r_{c}$=-1, $r_q$=-10 and the constant $C$ to $40$. The discount rate $\gamma$ is $0.95$. The stop threshold is $30$. The maximum dialogue length limit is set to $7$. The impact of $C$, and the stop threshold are shown in section \ref{different_environments}.

\section{Experiment Results}

\subsection{Offline Experiments}

\begin{table}
  \caption{Comparisons of CRM and the baselines on R (the average success rate), T (average number of turns), S (the success rate), W (the wrong quit rate) and L (the low rank rate).}
  \label{tab:dialogue_score}
  \begin{tabular}{ l | c | c | c | c | c }
    \hline
   	\hline
    Methods & R & T & S & W & L \\ \hline
    \hline
    MaxEnt@1 & -9.308 & 1.0 & 2.28 & 2.64 & 95.08 \\ \hline
    MaxEnt@2 & -6.496 & 1.523 & 12.60 & 5.44 & 81.96 \\ \hline
    MaxEnt@3 & 10.683 & 2.550 & 60.84 & 8.64 & 30.52 \\ \hline
    MaxEnt@4 & 20.648 & 3.492 & 82.08 & 8.60 & 9.32 \\ \hline
    MaxEnt Full & 20.670 & 4.351 & 84.36 & 8.68 & 6.96 \\ \hline
    CRM & 21.781 & 3.666 & 85.00 & 7.24 & 7.76 \\ \hline
  \end{tabular}
\end{table}

\subsubsection{Reinforcement Learning vs Greedy Methods}

First we explore how the reinforcement learning method helps the sequential decision making in the conversation system. Table~\ref{tab:dialogue_score} shows the experiment results of the Conversational Recommender Model (CRM) and the baselines on several key measurements. The result of CRM is generated by first training the RL agent for 20 epochs and selecting the best model, then evaluating on the test set. Comparing CRM and the MaxEnt Full model we observe that the RL agent finds a better policy. Specifically, the CRM agent is able to get the higher average reward in shorter average turns. Because the MaxEnt Full method always asks the facets one by one, until all facets are asked. However CRM learns that sometimes it is better to make a recommendation right away. This is critical in the conversation scenario. The users can easily get bored if they have to provide values for all the facets before seeing any recommendation. CRM also gets a higher success rate. This is also attributed to the fact that the RL agent has shorter turns. To be specific, it decreases the frequency of calling the belief tracker module. Due to the imperfection of the belief tracker model, more calls accumulates more errors, resulting in lower understanding accuracy. In other words, CRM decreases the possibility of misunderstanding caused by the belief tracker.

We further run several MaxEnt@K methods with $K<5$, and find that the reward increases as $K$ increases. We don't expect it to be generally true, especially if $K$ could be big, such as 100. However, in our data set, it seems $K=5$ (i.e. MaxEnt Full) is the best choice, thus we use it as the baseline for follow up experiments. 

We also evaluated the performance while not using the $s_t$ information, i.e. directly recommend items at the beginning of the dialogue. In this case, the model falls back to the plain FM model and performs extremely poor. Besides, we find that $s_t$ contributes to the candidate selection step in recommendation, however it does not seem to boost the FM model. 

The last two columns of Table~\ref{tab:dialogue_score} are also aligned with our expectation. When $k\geq  3$, CRM has a lower ``Wrong Quit Rate'' because it interacts with the NLU module less. When $k \leq 2$, the baselines have shorter ``Wrong Quit Rate'' because they interact with the belief tracker much less and perform poorly. For the ``Low Rank Rate'', CRM is slightly higher than the MaxEnt Full method because it is prone to make a faster recommendation and don't narrow the constraints as deep as the latter. Thus CRM's candidate list is often longer than the MaxEnt Full method, which makes it more challenging to rank the target item above the threshold. We have similar observations for other MaxEnt@K methods. We also examine the number of candidates in the last recommendation step for each mode. CRM generally has more candidates than the MaxEnt Full method, because MaxEnt Full adds more constrains than CRM.

\subsubsection{Recommendation Rewards}

\begin{table}
  \caption{Comparison of modeling the recommendation reward in different ways.}
  \label{tab:user_behavior}
  \begin{tabular}{ l || c | c | c | c | c | c }
    \hline
   	\hline
     & \multicolumn{3}{c|}{NDCG $r_p$} & \multicolumn{3}{c}{Cascade $r_p$} \\ \hline
    Model & R & T & S & R & T & S\\ \hline
    \hline
    MaxEnt Full & 10.799 & 4.351 & 84.24 & 10.783 & 4.351 & 84.20\\ \hline
    CRM & 11.772 & 3.830 & 85.80 & 11.537 & 3.804 & 84.88 \\ \hline
  \end{tabular}
\end{table}

To explore the impact of $r_p$, we run the experiments with the NDCG $r_p$ and the Cascade $r_p$ as discussed in Section \ref{recommendation_rewards}. We set $\alpha_1=\alpha_2=0.95$. Results are shown in Table~\ref{tab:user_behavior}. Note the Linear $r_p$ results have been listed in Table~\ref{tab:dialogue_score}. In all three settings, CRM always finds the better policy, with higher average reward and success rate in shorter turns. Linear $r_p$ and NDCG $r_p$ assume the user checks each item until threshold K, while Cascade $r_p$ assumes that the user may leave at any turn. We also observe that the reward of the Linear $r_p$ is higher than the other two. This is because NDCG $r_p$ and Cascade $r_p$ penalize the rewards nonlinearly with the decrease of the ranking. Changing the way of modeling $r_p$ is actually changing the environment for reinforcement learning. The experiment results reflect that CRM can consistently outperform the MaxEnt Full methods in different settings.

\subsubsection{The Impact of Belief Tracker Accuracy}

\begin{figure}[!ht]
\centering
\begin{minipage}{\linewidth}
\centering
\includegraphics[width=1\linewidth]{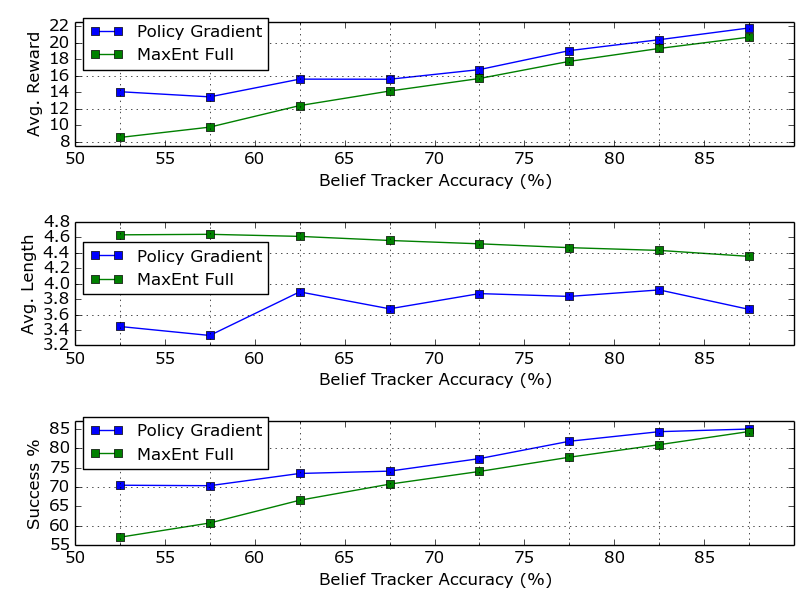}
\footnotesize
\caption{Model performances of three measures with different belief tracker accuracy.}
\label{fig:vary_belief_tracker}
\end{minipage}
\end{figure}

In our framework, we train the policy network based on the pretrained belief tracker. To study the impact of belief tracker accuracy, we explore how the two models perform on three major measures while varying the belief tracker accuracy from $52.5\%$ to $87.5\%$. 

Figure~\ref{fig:vary_belief_tracker} shows the belief tracker's accuracy has an important impact on the proposed framework. The Average Reward and the Success Rate increase as the belief tracker's error decreases. The Average Length of MaxEnt Full decreases a little as the accuracy grows, while CRM's length do not show obvious relationships with the belief tracker's performance. For all the cases, CRM is better than the baseline. Especially when the belief tracker's performance is poor. This reflects the robustness of the reinforcement learning model.

\subsubsection{Different Environments}
\label{different_environments}

\begin{figure}[!ht]
\centering
\begin{minipage}{\linewidth}
\centering
\includegraphics[width=1\linewidth]{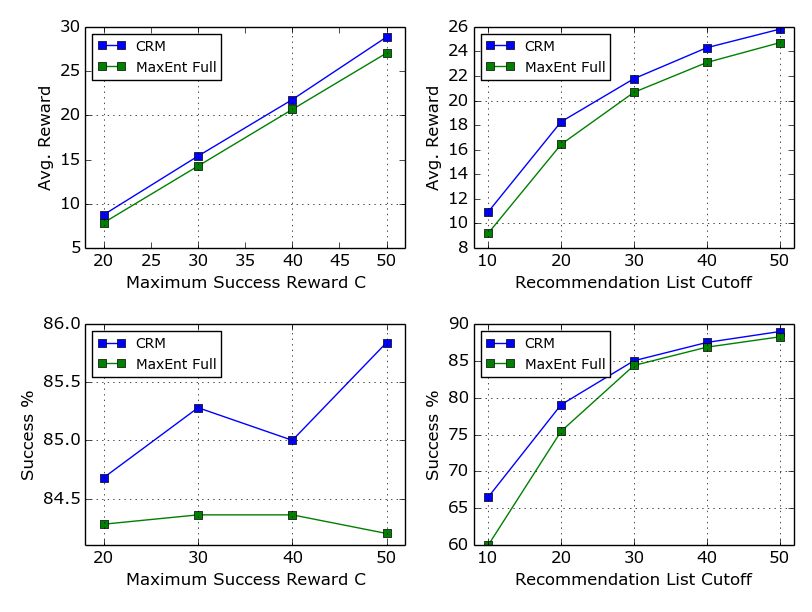}
\footnotesize
\caption{Comparison of CRM and the MaxEnt Full methods with different Maximum Success Reward C and stop threshold of the recommendation list.}
\label{fig:vary_environment}
\end{minipage}
\end{figure}

To study the effects of the simulated environments, we vary two major factors, the Maximum Success Reward $C$ and the Recommendation List Stop Threshold. Results are shown in Figure~\ref{fig:vary_environment}. The left two plots show how the measures varies with $C$. The right two plots are for the threshold. We observe an increase for the Average Reward and the Success Rate as $C$ grows or the stop threshold increases. In all cases, CRM outperforms the baseline.

With further analysis, we found that the average conversational length of CRM increases with $C$. This is not surprising, because as $C$ increases, it does worth the effort for the RL agent to spend more time to gather information so as to increase the chance of recommending the target item, thus receiving the reward.

\subsection{Online User Study}

We further evaluate our trained model with the online crowd sourcing experiments and present the quantitative results here. The ideal users would be those yelp user who have actually visited a number of restaurants and would like to chat with our agent to inform her current interest of a target. And our agent would make a recommendation at the end based on her historical interests as well as the current session's user intention. However, it is relatively hard for us to find those real users. Instead we come up with an experimental design to try to recover the ideal scenario as much as possible.

First we randomly sample a target restaurant from the test dataset, containing a user id, a restaurant id, and the facets of this restaurant. Second, using the sampled user id, we retrieve a list of the restaurants that are visited by this user from the train set. This list is treated as the ``historical'' visiting information. Next, an Amazon Mechanical Turk worker is presented with the list of visited restaurants with their meta data. They are instructed to view each item carefully, in order to ``learn the preference'' of the sampled user. After this, the worker begins to chat with the conversational recommendation agent. The worker is presented with the facet values of the target restaurant on the side, so that she can correctly answer the questions. However, she doesn't know which restaurant is the target. The agent may fail due to the error in the belief tracker module. At the end, the worker needs to select up to three restaurants in the recommendation list. To motivate the worker to work carefully, she would receive a bonus if she successfully finds the target restaurant. Unlike the User Utterance Collection phase for collecting simulated offline data, there is no special constraints in this online study and the users can chat freely in natural language and quit a session as they want. 

The MaxEnt Full method based agent is used as a baseline, since it's better than other $MaxEnt@*$ models. Both of the two agents use the same belief tracker and recommender. Each worker needs to complete 8 dialogues. A worker interacts with our two agents in a random order, and each subject wouldn't experiment the same target restaurant more than once. 

We collected 208 dialogues, with 104 dialogues for each method. The linear $r_p$ recommendation reward is used. We set $C=40$ and the stop threshold to $30$. A dialogue is successful only if the worker actually found the target restaurant from the recommendation list. 

The results in Table~\ref{tab:user_study_result} show that in general CRM is better, with a higher reward of $0.7371$. Note that the average reward and the success rate are lower than the offline experiments, indicating that the real online settings is more complicated than the offline settings. CRM achieves a higher success rate of 28.846\%, showing that it can better assist the real users to find the target items. Specifically, $30$ out of $104$ users successfully find their target items when using CRM, versus 23 out of 104 for the baseline. CRM has in a shorter turn of 3.79 compared to 4.58 of the baseline. This is aligned with the offline experiments, because CRM does not collect all the facets before making a recommendation.

We further explored the average ranking of the target item in the recommendation list, and that of the policy gradient model is 3.33 and that of the baseline is 1.91. The reason for this is because the MaxEnt Full model often has a shorter candidate list because it always requests the values of all the facets. In future we plan to improve this aspect by introducing a ranking oriented recommender algorithm to our framework. 
Table~\ref{tab:dialogue_sample} shows the example dialogues collected in the user study. The MaxEnt Full method asks all slots before recommendation, while CRM often recommends earlier after requesting only 3 or 4 facets.

\begin{table}
  \caption{Comparison of CRM and the MaxEnt Full methods in online user study.}
  \label{tab:user_study_result}
  \begin{tabular}{ l | c | c | c | c | c }
    \hline
   	\hline
    Model & R & T & S \\ \hline
    \hline
    MaxEnt Full & -2.7884 & 4.58 & 22.115 \\ \hline
    CRM & 0.7371 & 3.79 & 28.846 \\ \hline
  \end{tabular}
\end{table}

\begin{table*}
\small
  \caption{Sample dialogues of CRM and the MaxEnt Full methods. The facet values extracted by the NLU module are shown in the parentheses.}
  \label{tab:dialogue_sample}
  \begin{tabular}{ l | l }
    \hline
   	\hline
    CRM & MaxEnt Full \\ \hline
    \hline
    \makecell{
      \underline{User}: I NEED MEXICAN CATEGORY \hspace{0.35cm}\textit{(category="Mexican")}\\
      \underline{Agent}: Which city are you in?\\
      \underline{User}: GILBERT \hspace{3.5cm} \textit{(city="Gilbert")}\\
      \underline{Agent}: Which rating range do you want?\\
      \underline{User}: 4.0 \hspace{3.55cm} \textit{(rating_range>="4.0")}\\
      \underline{Agent}: <make recommendation, target ranked 7 of 12>\\
      } & 
      \makecell{
      \underline{User}: Find me some Mexican restaurants in AZ \\
      \hspace{3.35cm} \textit{(category="Mexican", state="AZ")}\\
      \underline{Agent}: Which city are you in? \\
      \underline{User}: Gilbert \hspace{3.8cm} \textit{(city="Gilbert")}\\
      \underline{Agent}: Which rating range do you want?\\
      \underline{User}: 4 \hspace{3.75cm} \textit{(rating_range>="4.0")}\\
      \underline{Agent}: Which price level would you like?\\
      \underline{User}: cheap pricing. \hspace{2.25cm} \textit{(price_range="cheap")}\\
      \underline{Agent}: <make recommendation, target ranked 7 of 12>\\
      }\\ \hline
      Note: The RL agent asked for three slots and then make & Note: The rule agent collected all five slots before\\
     an recommendation, no need to ask for the rests. &  recommendation \\
    \hline
      
    \makecell{
      \underline{User}: juice bars \& smoothies\\
      \hspace{3cm} \textit{(category="Juice Bars \& Smoothies")}\\
      \underline{Agent}: Which city are you in?\\
      \underline{User}: henderson \hspace{3cm} \textit{(city="Henderson")}\\
      \underline{Agent}: Which rating range do you like?\\
      \underline{User}: 4.0 \hspace{3.55cm} \textit{(rating_range>="4.0")}\\
      \underline{Agent}: <make recommendation, target ranked 2 of 4>\\
      } & 
      \makecell{
      \underline{User}: I NEED BEST Juice Bars \& Smoothies \\
      \hspace{3cm} \textit{(category="Juice Bars \& Smoothies")}\\
      \underline{Agent}: Which city are you in?\\
      \underline{User}: IN hENDERSON \hspace{2.3cm} \textit{(city="Henderson")}\\
      \underline{Agent}: Which rating range do you like?\\
      \underline{User}: 4.0 \hspace{3.55cm} \textit{(rating_range>="4.0")}\\
      \underline{Agent}: Which price range do you want?\\
      \underline{User}: MEDIUM PRICE \hspace{1.63cm} \textit{(price_range="medium")}\\
      \underline{Agent}: <make recommendation, target ranked 1 of 1> \\
      }\\ \hline
    Note: three slots are asked by the agent 
    & 
    Note: four slots are collected by the agent \;\;\;\;\;\\ \hline
    \hline
  \end{tabular}
\end{table*}

\section{Conclusion and Future Work}

We argue and propose a unified framework to integrate recommender systems and dialogue system technologies together for building an intelligent conversational recommender system. We tailored the important building blocks of dialogue systems for the tasks of the conversational recommendation agents. Under this framework, machine action space and user states are clearly defined. The state of the bot (i.e. the belief of a user's information need or the user query) is represented and constantly updated as semi-structured data as the conversational agent communicates with a user and gathers more information from the user. 

Instead of a greedy approach that just returns the top ranking results given the current user query, the agent takes actions and presents information to optimize for long term reward such as a higher successful rate, a shorter turn or delayed reward. We introduce the reward function for the conversational system based on the recommendation systems research. The dialogue agent can learn which action maximizes the session based reward at each step with reinforcement learning. It learns to collect the facet values as needed and make a recommendation directly when appropriately. We have developed a demo system. Both the online and offline evaluation results demonstrate the merits of introducing recommendation techniques into a reinforcement learning based conversational system.  

This work is a first step towards conversational recommendation agents. The work has much limitation and much room for further improvement. For example, we can have a unified model that jointly learns the dialogue policy and the recommendation model at the same time. We can also improve the facet search components and improve the design and evaluation of our framework based on these models. Third, for the belief tracker, we train it by first collecting user utterances, converting them to delexicalized templates, and then simulating more dialogues using the templates. Although simulating user with simple schedule with strong assumptions is a common practice in both reinforcement learning studies and the dialogue community, as the model trained with simulation can be a reasonable starting point, we would like to further improve the work with better simulated users and less assumptions, or with online learning while chatting with real users. Fourth, the study only includes two types of actions: requesting a facet value from users and making recommendations to the users. The machine action space could be much bigger. For example, users can ask questions proactively or machine can ask feedback on particular items, ask for confirmation, let user change mind, etc. 
Last but not least, we can also explore different reward functions, such as multi session rewards, life time value of a user or revenue, etc.

\bibliographystyle{ACM-Reference-Format}
\balance
\bibliography{sample-bibliography} 

\end{document}